\documentclass[preprint2,twocolumn,times,tighten]{aastex62}
\usepackage{color}
\usepackage{enumitem}
\usepackage{amsmath,amstext,mathrsfs}

\usepackage{multirow}
\newcommand{\specialcell}[2][c]{\begin{tabular}[#1]{@{}c@{}}#2\end{tabular}}

\shorttitle{Comment on Clark et al, 2019}
\shortauthors{Yuen et al.}

\begin{document}
\title{Comment on Clark et al. (2019) "The Physical Nature of Neutral Hydrogen Intensity Structure"}
\correspondingauthor{Ka Ho Yuen}
\email{kyuen2@wisc.edu}

\author[0000-0002-0786-7307]{Ka Ho Yuen} 
\affiliation{Department of Astronomy, University of Wisconsin-Madison}

\author{Yue Hu}
\affiliation{Department of Physics, University of Wisconsin-Madison}
\affiliation{Department of Astronomy, University of Wisconsin-Madison}

\author{A. Lazarian}
\affiliation{Department of Astronomy, University of Wisconsin-Madison}

\author{D. Pogosyan}
\affiliation{Department of Physics, University of Alberta, Canada}

\begin{abstract}
A recent arXiv publication by \citet{susan19} (CX19) uses both GALFA-\ion{H}{1} observational data and numerical simulations to address the nature of intensity fluctuations in Position-Position-Velocity (PPV) space. The study questions the validity and applicability of the statistical theory of PPV space fluctuations formulated in \citet{LP00} (LP00) to \ion{H}{1} gas and concludes that $"${\it a significant reassessment of many observational and  theoretical studies of turbulence in \ion{H}{1}}$"$. This implies that dozens of papers that used LP00 theory to explore interstellar turbulence as well as the ongoing research based on LP00 theory are in error. This situation motivates the urgency of our public response.  In our {\it Comment} we explain why we believe the criticism in CX19 is based on the incorrect understanding of the LP00 theory. We reveal problems with the assumptions, analysis and the interpretations in CX19. In particular, we illustrate that the computation of correlation between PPV slices and dust emissions in CX19 does not properly reveal the relative importance of velocity and density fluctuations in velocity channel maps. In fact, we show that the reported absence of velocity contributions in PPV slices follows directly from the LP00 theory, and illustrate that even for "velocity caustics" one can obtain a relatively high correlation to the synthetic dust intensity map. While CX19 provides its explanation of the change of the spectral index with the change of the thickness of the PPV slice that is based on the two phase nature of \ion{H}{1} gas, we failed to see any observational support for this idea. On the contrary, we show that the observations both in two phase \ion{H}{1} and one phase CO show similar results. Moreover, the observed change is in good agreement with LP00 predictions and spectral indexes of velocity and density spectra that are obtained following LP00 procedures are in good agreement with the numerically confirmed expectations of compressible MHD turbulence theory. We also address the arguments about origin of density filaments in CX19 that are based on 128$^3$ MHD turbulence simulations. Increasing the resolution to 1200$^3$ we come to the conclusion that is opposite to that in CX19, namely, that a significant part of the structures in thin PPV slices arises from velocity caustics even at high sonic Mach number. In short, we could not find any justification of the criticism of LP00 theory that is provided in CX19. On the contrary, our analysis testifies that both available observational and numerical data agree well with the predictions of LP00 theory. 
\end{abstract}


\section{Introduction} \label{sec:intro}

Spectroscopic Doppler-shifted lines carry information about astrophysical turbulence. The corresponding data for \ion{H}{1} and molecular lines, e.g. CO is stored in Position-Position-Velocity (PPV) data cubes. The theory that relates the statistics of spectroscopic intensity fluctuations in the PPV space and the underlying statistics of turbulent velocities and densities was developed in \citet{LP00} (henceforth LP00) and extended in the subsequent theoretical studies\citep{LP04,LP06,LP08,KLP16,KLP17a,KLP17b}. The theory was extensively tested with numerical data in \citep{2008arXiv0811.0845C} and was applied to numerous sets of observational data of \ion{H}{1} and CO data. Together with the modern theories of MHD turbulence \citep{GS95} and turbulent reconnection \citep{LV99}, LP00 theory is at the foundations of a new Velocity Channel Gradient (VChG) technique \citep{LY18a} that has been shown to successfully trace magnetic fields both in \ion{H}{1} and molecular clouds \citep{2018MNRAS.480.1333H,survey}.

A recent work by \citet{susan19} (henceforth CX19) challenged the results obtained on the basis of LP00. The study analyses quantitatively on \ion{H}{1} data and qualitatively in numerical simulations and come to the conclusion that  "{\it a significant reassessment of many observational and theoretical studies of turbulence in \ion{H}{1}" is necessary.}" In view of this strong claim, in the Comment below we provide analytic, numerical and observational arguments testifying that the conclusions in CX19 based on the mis-interpretation of the LP00 theory. In future we are going to summarize our arguments in fully fledged paper, but to avoid uncontrolled spreading of the confusion about the nature of LP00 findings we use the form of a "Comment" instead. 

\begin{figure*}
\centering
\includegraphics[width=0.99\linewidth]{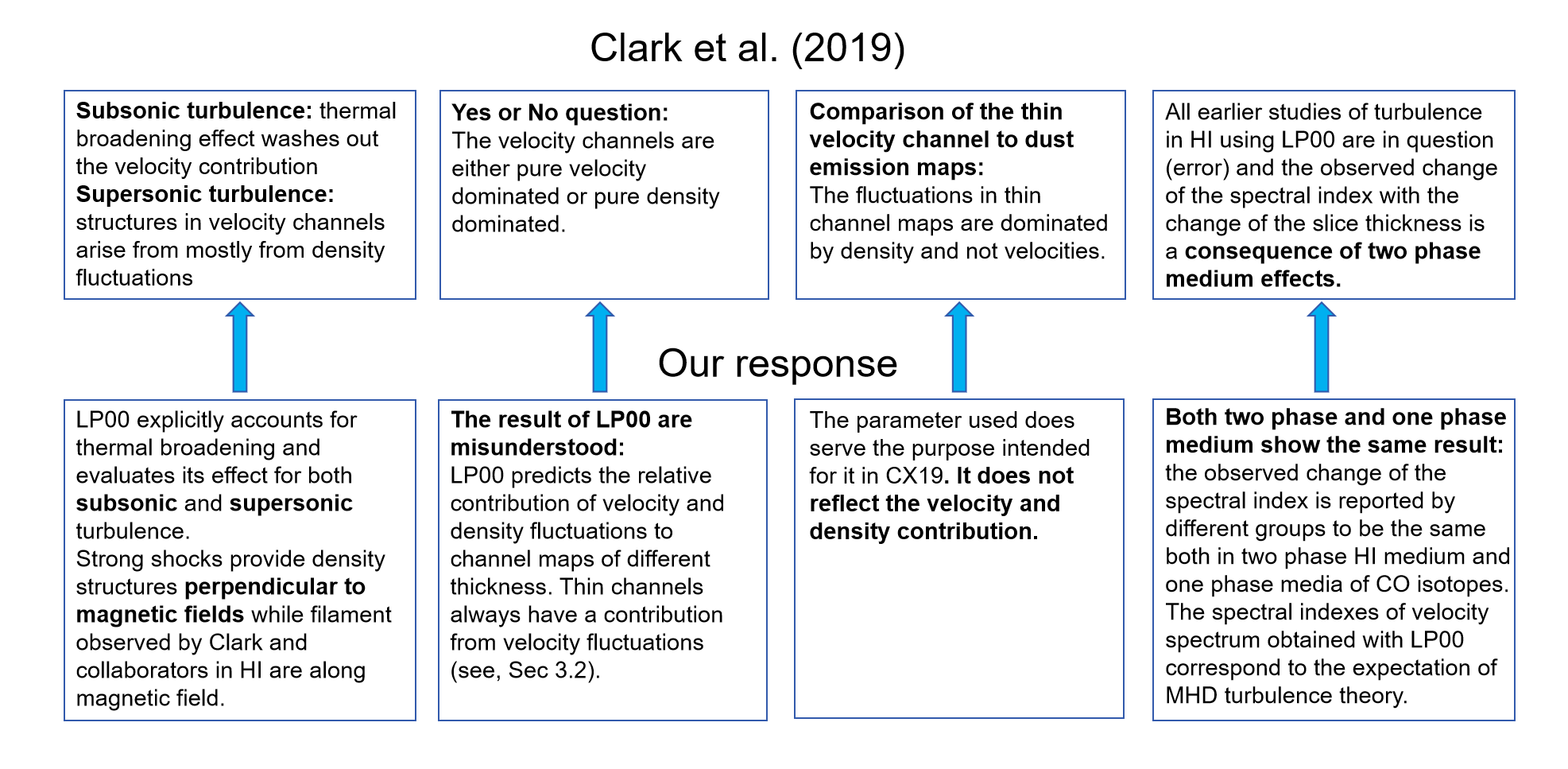}
\caption{Summary of points in \citet{susan19} and summary of our response.}
\label{fig:summary}
\end{figure*}

Fig.\ref{fig:summary} provides a summary of main points of CX19 as well as our concise response to these points. In what follows we provide an extended explanation why we do not agree with the criticism of LP00 theory that is provided in CX19.

\section{LP00: Both density and velocity form structures }
\label{sec:LP00}

One of the central statements that CX19 made is from the 1st paragraph of section 3.2: {\it ``If the \ion{H}{1} intensity structures are velocity caustics, there should be no enhancement in the FIR emission strength at the location of the intensity structures, relative to the surrounding medium. Conversely, an enhancement in the FIR emission indicates that the structures are true density features.''}. 
This statement reflects a false dichotomy permeating CX19 that the structures in channel maps are either velocity caustic or true density features.  Such dichotomy is incongruous with the theoretical picture advanced in LP00.

To clarify why we claim that velocity effects are important for intensity structures observed in thin velocity channels, we have to recall the foundations of LP00 theory. 

The density in PPV space of emitters with temperature $\beta_T$ moving along the line-of-sight with stochastic turbulent velocity $u(\bf x)$ and regular coherent velocity $v_{\mathrm{g}}(\bf x)$ is
\begin{equation}
\rho_s(\mathbf{X},v) =\int_0^S \!\!\! dz 
\frac{\rho(\mathbf{x})}{\sqrt{2\pi \beta_T}}
\exp\left[-\frac{(v-v_{\mathrm{g}}(\mathbf{x}) -
u({\bf x}))^2} {2 \beta_T }\right] 
\label{eq:rho_PPV}
\end{equation}
where sky position is described by 2D vector $\mathbf{X}=(x,y)$ and $z$ is the line-of-sight coordinate. The Eq.~(\ref{eq:rho_PPV}) is \textit{exact}, including the case when the temperature of emitters varies in space, $\beta_T = \beta_T(\mathbf{x})$. 

The Eq.~(\ref{eq:rho_PPV}) represents the effect of the velocity-dependent mapping from the Position-Position-Position (PPP) space to PPV space. 
Due to this mapping, the PPV density $\rho_s(\mathbf{X},v)$ at a given velocity $v$ is determined both by the spatial density of the emitters $\rho(x,y,z)$ and their respective line-of-sight velocities.  Note that formal caustics, understood as singularities of differentiable map \cite{arnold1985singularities} from PPP to PPV space, arise only in the limit of $\beta_T \to 0$.  For finite temperatures it is better to talk about ``velocity crowding''  in the PPV space. This velocity crowding effects result in {\it modified} intensity enhancements in channel maps which leads to modifications in the power spectrum of velocity channels.

In observations, the intensity of the velocity channel with a center velocity $v_c$ and a finite  width $\Delta v$ is related
to PPV density by the convolution with a channel window $W_{\Delta v}(v)$: 
\begin{equation}
\rho_s'(\mathbf{X},v_c,\Delta v) =\int \!\! dv \; W_{\Delta v}(v_c-v) \rho_s(\mathbf{X},v)
\end{equation}
If channel integration does not extend over the whole line profile, velocity effects continue to contribute to the channel. It is a matter whether the contribution of velocity is important to account for or not in the particular analysis, not an issue of existence of velocity effects. LP00 teaches us how and when density and velocity effects can be \textit{statistically} disentangled. To help our reader to understand the main concepts in LP00, in Appendix \ref{apsec:A} we provide a summary of the results from LP00 and the subsequent papers, which were later called the Velocity Channel Analysis (VCA).

\label{sec:ncc}

\section{What does the parameter $\Delta I_{857}$ tell us?}
\label{sec:result}

CX19 has introduced the parameter:
\begin{equation}
    \Delta I_{857}=\frac{\sum_{i=1}^{n}I_{i}*\omega_{i}^{\Delta v}}{\sum_{i=1}^{n}\omega_{i}^{\Delta v}}-\frac{\sum_{i=1}^{n}I_{i}}{\sum_{i=1}^{n}n}
    \label{eq:susan_parameter}
\end{equation}
The parameter $\Delta I_{857}$ constitutes the difference between intensity $I_{857}$  of Planck $857~GHz$ dust emission map \textit{weighted with the GALFA high-pass Un-Sharp Mask (USM) filtered channel intensity $w^{\Delta v}$} and the mean of pure $I_{857}$. CX19 has used its non-zero value and flat dependence on channel thickness to argue that even thin velocity channels are dominated by real space density structures, and, moreover, claim
the failure of theoretical velocity channel analysis based on LP00.

In what follows we first summarize the points made by CX19 in \S 3.1 and provide our response. We then demonstrate that the measure $\Delta I_{857}$ introduced in CX19 does not serve the purpose intended for it by CX19 (\S 3.2). In \S 3.3 we comment on the CX19 measurements of variability of 
$I_{857}/\omega$ as the function of $\omega$.

\subsection{Methodology issues of $\Delta I_{857}$ in CX19}
We summarize the arguments in CX19 related to the $\Delta I_{857}$ parameter
\begin{enumerate}
\item CX19 claimed that one should expect the parameter to be near zero for sufficiently thin channels, to quote "{\it If the \ion{H}{1} intensity structure is caused entirely by velocity caustics, $\omega_{i}^{\Delta v}$ will be uncorrelated with the density field, and $\Delta I_{857}$ will equal 0}.  while at larger channel width the parameter will grow {\it linearly}. This wedge-like behaviour is qualitatively represented in lower panel Fig.~\ref{fig:GALFA228} which we extract from CX19 here for clarity. 
\begin{figure*}
\centering
\includegraphics[width=0.98\linewidth,height=1.2\linewidth]{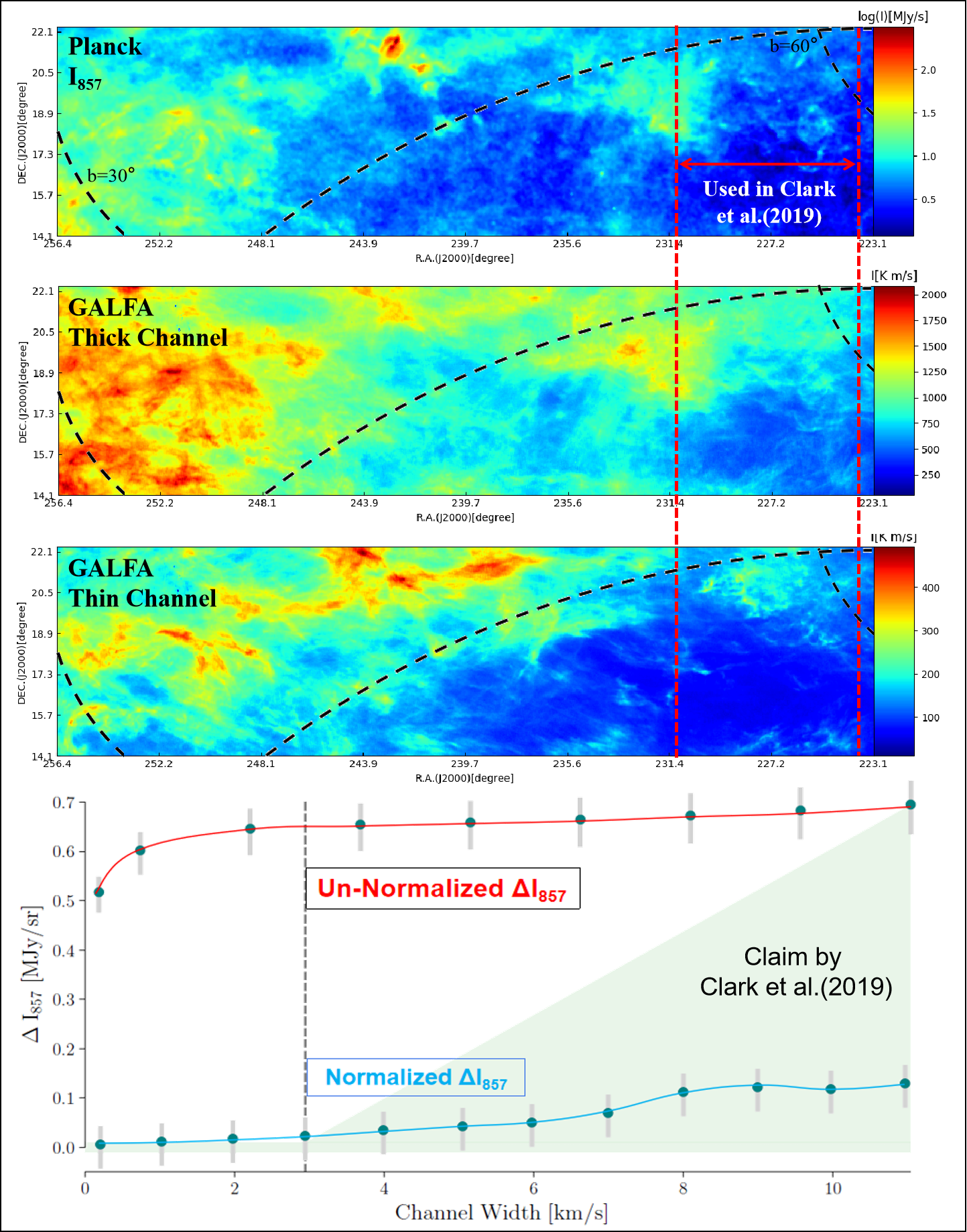}
\caption{Planck 857 GHz (top panel) and GALFA-\ion{H}{1} channel intensity maps (two middle panels) with channel width $\Delta v = 42 km/s$ (thick), and channel width $\Delta v = 2.2 km/s$ (thin). The red dashed line indicates the region used in \citep{susan19}. Bottom panel: the plot of un-normalized $\Delta I_{857}$,  as a function of the velocity channel width , reproduced from \citet{susan19}. The pale green wedge represents  unjustified theoretical prediction as claimed by \citet{susan19}. Gray dashed line represents $\Delta v = 2.94 km s^{-1}$, the velocity channel width used in \citet{LY18a}. Blue line shows normalized cross-correlation coefficient $\widehat{\Delta I_{857}}$ that measures similarity of
$I_{857}$ and USM filtered GALFA-\ion{H}{1} channel map. Note that with
some abuse of the notation, we used
the same y-axis to mark the magnitudes of both dimensional $\Delta I_{857}$ and dimensionless $\widehat{\Delta I_{857}}$ }
\label{fig:GALFA228}
\end{figure*}

\item Based on this, CX19 argue that the measured $\Delta I_{857}$ which has high value compared to the green wedge (See the lower panel of Fig.\ref{fig:GALFA228}), and is practically constant across all channel widths, is inconsistent with theoretical expectation from LP00.

\item On the basis of the above, CX19 concludes that the structures in thin GALFA velocity channels cannot be caused by velocity effects, and therefore reflect density fluctuations.
\end{enumerate}

We shall demonstrate that all three steps are faulty, namely
\begin{enumerate}

\item We show that VCA theory (see \S \ref{apsec:A}) under LP00 assumptions, predicts $\Delta I_{857}$ parameter to be
independent on the channel thickness. \textit{Wedge-like expectation with special suppression for thin channels does not follow from the theory of LP00 and misrepresents LP00 results}. Comparison of any measured data to this green-wedge "prediction" therefore did not bring any conclusions to what the CX19 claimed. 

\item We show that the theory predicts a non-zero value for $\Delta I_{857}$ as long as density inhomogeneities are present and this
value does not reflect velocity effects. The reason is that $\Delta I_{857}$ measure projects out the effect
of any turbulent velocities that are uncorrelated to density.\textit{The measurements of $\Delta I_{857}$ in CX19 do match the true predictions of LP00 theory.} 

\item Since $\Delta I_{857}$ projects out velocity effects of PPV mapping, \textit{$\Delta I_{857}$ parameter cannot be used to judge the relative level of velocity contribution.}
\end{enumerate}

\subsection{Theoretical point of view: What does the correlation between thin channel maps and intensity map tells us?}
\label{ssec:theory}

The Planck $857~GHz$ dust emission is dominated by thermal dust and its intensity is proportional to dust column density, which is expected to follow \ion{H}{1} density. Thus $\Delta I_{857}$ parameter,
which can be rewritten as
\begin{align}
    \Delta I_{857} = \frac{\langle I\omega\rangle - \langle I\rangle\langle \omega\rangle}{\langle \omega \rangle}~,
\end{align}
essentially measures the cross correlation between column density and channel density fluctuations in PPV cube.
The cross-correlation is normalized by the mean USM-filtered density in a channel, though a useful additional normalization by the mean $I_{857}$
is missing, which adds difficulty in making conclusions based on the  parameter magnitude.

So let us use the framework of LP00 to see what one can learn about velocity relative contributions by measuring a parameter similar to $\Delta I_{857}$. Consider two signals, one, $I$, proportional to the local column density, $ I(\mathbf{X}_1)  \propto \int dv \rho_s(\mathbf{X}_1,v)
\equiv \rho_c(\mathbf{X}_1)$ and another, $w$, proportional 
to PPV channel density, $ w(\mathbf{X}_2)  \propto \int^{\Delta v} dv \rho_s(\mathbf{X}_2,v) \equiv \rho_s^{\Delta v}(\mathbf{X}_2,v)$.
Given that column density has no velocity information, it is no surprise that correlation between the two \textit{does not furnish information about turbulent velocities at all, if turbulent velocities and density, as assumed in LP00, are uncorrelated}. Indeed, as it is not difficult to see from, for instance, Appendix~B of \cite{LP04}, the velocity part just factorizes as a mean channel density from such cross correlations
\begin{equation}
\left\langle \rho_c(\mathbf{X_1}) \rho_s^{\Delta v}(\mathbf{X}_2,v) \right\rangle \approx
\frac{\left\langle \rho_s^{\Delta v}(v)\right\rangle}
{\left\langle \rho_c \right\rangle}
\left\langle \rho_c(\mathbf{X_1}) \rho_c(\mathbf{X_2}) \right\rangle
\label{eq:5}
\end{equation}
As the result, the cross correlation of our signals
\begin{equation}
\left\langle \frac{I(\mathbf{X_1})}{
\left\langle I \right\rangle} 
\frac{w(\mathbf{X}_2,v,\Delta v)}{\left\langle w(v,\Delta v)\right\rangle} \right\rangle
\approx 
\frac{\left\langle \rho_c(\mathbf{X_1}) \rho_c(\mathbf{X_2}) \right\rangle}
{\langle \rho_c \rangle^{2}}
\label{eq:Iw_corr}
\end{equation}
is not sensitive to the velocity contribution at all. This conclusion holds also when linear filtering on the sky is applied to the signals.

To summarize, the LP00 prediction for $\Delta I_{857}$ parameter is
\begin{equation}
    \Delta I_{857} \approx \langle I_{857} \rangle \left( \frac{\left\langle \rho_c(\mathbf{X_1}) \rho_c(\mathbf{X_2}) \right\rangle} {\langle \rho_c \rangle^{2}}  -  1\right)
\end{equation}
which is column intensity dependent. Essentially, $\Delta I_{857}$ projects out any velocity contribution that is not correlated to density. Velocity effects can be arbitrarily large and define channel intensity non-uniformity, but will not enter the measure directly. \footnote{Note that LP00 has used the assumption of negligible correlation between density and turbulent velocities in a specific narrow mathematical sense --- that correlation function between density and $z$-component of turbulent velocity $\xi_{\rho u_z}(R,z) \equiv \langle \rho(\mathbf{x}) u_z(\mathbf{x+r})\rangle$ can be approximately neglected while computing correlation properties of channel intensities, as opposite to either looser ideas that 'density causes velocity' or to correlations of density with velocity derivatives, especially divergence, that is often present. This has some theoretical basis in the properties of $\xi_{\rho u_z}$ (it is an odd function  $\xi_{\rho u_z}(R,z)=-\xi_{\rho u_z}(R,-z)$ and enters double integral along pair of lines of sight equally with positive and negative sign) but ultimately depends on the properties of the turbulence and require validation in simulations.}

In Figure~\ref{fig:GALFA228} we also plot our measurements of the normalized cross-correlation coefficient between FIR and USM filtered channel maps, $\widehat{\Delta I_{858}} = \Delta I_{857} \times \langle \omega \rangle / \langle I \rangle$, i.e cross-correlation normalized by the mean intensities. This parameter does not depend on the intensities and thus the number carries the meaning of how the enhancements of densities are compared to the mean intensities. As we see, correlation between two maps is weak for any channel width, which is not of course surprising, since the two maps are filtered differently, and consist of different modes. Some dependence on the channel width comes from the channel intensity variances.

\subsection{The possibility of "Cold Neutral Media"?}

CX19 investigated the spatial variation of $I_{857}/N_{HI}$ ratio \citep{2015A&A...577A.110Y,2018ApJ...862...49N}
and found that it is enhanced  in the regions with higher value
of USM filtered channel intensity $\omega^{\delta v}$.
CX19 interpreted this finding as an argument that high intensity
small scale channel structures are dominated by Cold Neutral Medium (CNM) and are physically distinct from surrounding medium.

\begin{figure}
\centering
\includegraphics[width=0.98\linewidth,height=1.0\linewidth]{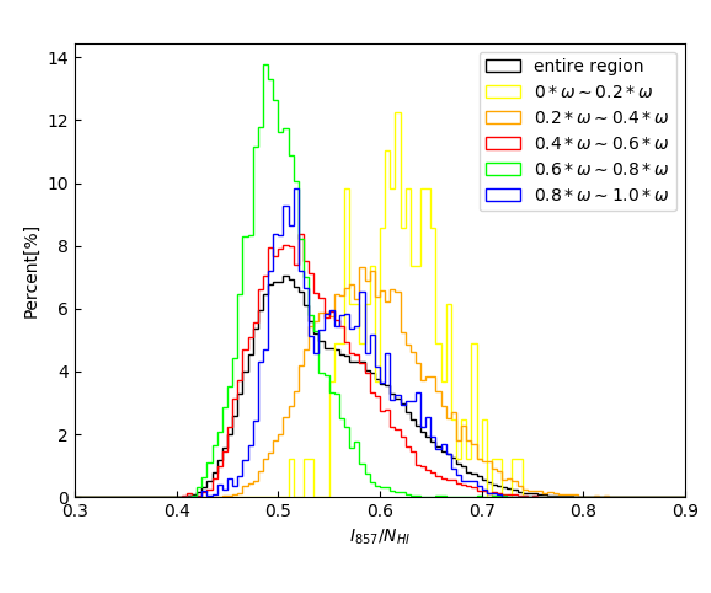}
\caption{Counter example to CX19 using the GALFA-\ion{H}{1} data of region Z, spanning Right Ascension (R.A.) 225.2$^\circ$ to 232.2$^\circ$ and Declination (DEC.) 14.1$^\circ$ to 22.7$^\circ$.  Histogram of $I_{857}/N_{HI}$ as a function of USM intensity shows the dependence on the small scale channel intensity very different from that shown in CX19 in their Fig.~8.  We analysed the \ion{H}{1} data from -21$km/s$ to 21$km/s$, thin channel map with channel width $\Delta v=2.2 km/s$ and the column density data integrated from -90$km/s$ to 90$km/s$ .
}
\label{fig:NHI}
\end{figure}

CX19 results are presented in Figure 8 of their paper as the histogram of $I_{857}/N_{HI}$ ratio drawn from extended region of the GALFA-\ion{H}{1} sky with $|b| > 30^{\circ}|$ and $N_{HI} < 8\times10^{20}cm^{-2}$. However, our measurements have shown that the tendency to have higher $I_{857}/N_{HI}$ at positions of velocity channel structures are not universally replicated when analysis is done over smaller specific sky patches. For instance, Fig.~\ref{fig:NHI}, that shows analysis for the region shown in Fig.~\ref{fig:GALFA228}, demonstrates practically an opposite trend. The results of CX19 are reproduced only when sky coverage is extended. This does not lend support to the conclusion that channel structures have physical relation to CNM,
when one would expect object by object enhancement of $I_{857}/N_{HI}$, but points to more a statistical effect. In addition, the location of sample region in Fig.~8 of CX19 is close to the North Galactic Pole, showing in particular high contrast channel structures up to $b \sim 87^\circ$ where neither formation of cold \ion{H}{1} nor an increase of dust emissivity \citep{2016A&A...586A.138P} is likely. 

\section{Do observational results agree with LP00 or CX19 expectations?}
\label{subsec:4.1}

CX19 questions the validity of the results obtained in the last 20 years by different groups who used LP00 theory to analyze their data. This is a very serious claim. The most valuable insight from LP00 is the prediction of the spectral slope change between the thin and thick PPV slices that is related to the spectral indexes of turbulent velocity and density as it is demonstrated in Tab.~\ref{tab:L09}.\footnote{LP00 prediction is radically different from the one that can be obtained assuming that we measure the spectrum of the thin and thick {\it density} cubes. Assume that it is the regular shear maps the fluctuations of density into the PPV space, while the effects of turbulent velocity on the PPV statistics are suppressed (which broadly corresponds to the message in CX19). Then the spectral index in thin and thick trivially should differ by unity, contrary to what is seen in observations.} The corresponding LP00 technique is termed Velocity Channel Analysis (VCA). 

Note that a different technique, the Velocity Coordinate Spectrum (VCS), was also suggested in LP00 (see also \citet{2006astro.ph.11465C,LP06}). The VCS expressions obtained in LP00 relate the spectral index of PPV intensity fluctuations along the V-coordinate and the underlying velocity spectrum (see Table.\ref{tab:VCS}). The VCA and the VCS were used by different groups with some of the results summarized in Tab.~\ref{tab:4} and Tab.~\ref{tab:5}. These results also testify the importance of velocity fluctuations in thin slices and the obtained spectra also corresponds to the expectation of compressible MHD theory.

CX19 provides a different explanation of change of the spectral index within the thin and thick velocity slices in \ion{H}{1}. They do not provide any quantitative predictions for the change\footnote{We have serious problems to understand on a qualitative level why this change can take place. Indeed, it is not clear why sampling two phase gas with can produce anything but a change of the spectrum by unity, i.e. from $K^{n+1}$ in a thin slice to $K^{n}$ in a volume. However, this change is not related to gas having two phases. This is also contradicts to observations. See also \S \ref{subsec:4.3}}, but speculate that the change of the spectral index can be due to the two phase nature of the emitting media. This interpretation suggests that the correspondence of the velocity spectral index $E_v^A$ measured using the VCA and spectra $E_v^C$, $E_v^S$ as well as the correspondence of the $E_v^A$ and $E_\rho$ to the expectations of the MHD turbulence theory is purely coincidental, the claim that we find to be highly improbable. Moreover, CX19 interpretation suggests that differences are expected in the studies of spectra of the two phase \ion{H}{1} and single phase CO.  
 
From Tab.~\ref{tab:4}, we can see the observed change of the spectral index is reported by different groups to be the same both in two phase \ion{H}{1} medium and one phase media of CO isotopes. Therefore, this is another evidence that the change of the spectral index cannot be a consequence of the two-phase media effects.

\begin{table*}
\centering
\label{tab:L09}
\begin{tabular}{| c | c | c |}
\hline
Slice thickness & \specialcell{Shallow 3D density spectrum \\
$ E_\rho \propto k^{-1+\gamma}$, $\gamma >0$} & 
\specialcell{Steep 3D density spectrum\\
$E_{\rho} \propto k^{-1+\gamma}$, $\gamma<0$} \\ \hline \hline
Thin slice   & $P_{2D} \propto K^{-3+\gamma+m/2}$ & $P_{2D}  \propto K^{-3+m/2}$ \\\hline
Very Thick slice  & $P_{2D} \propto K^{-3+\gamma}$ & $P_{2D} \propto K^{-3+\gamma}$\\
\hline
\end{tabular}
\caption{The spectrum asymptotics of velocity channels in the limiting cases studied in LP00. $\gamma$ is the reduced density power spectrum slope while $m$ is the velocity structure function index $D_z \sim \langle u_z({\bf r'}) - u_z({\bf r}+{\bf r'}) \rangle \propto r^m$, which corresponds to velocity energy spectrum
$E_v \propto k^{-1-m}$. }
\end{table*}

\begin{table*}
\centering
\label{tab:VCS}
\begin{tabular}{| c | c | c |}
\hline
Spectral term & $\Delta B < S[k_v^2D_z(S)]^{-\frac{1}{m}}$ & $\Delta B > S[k_v^2D_z(S)]^{-\frac{1}{m}}$ \\ \hline \hline
$P_{\rho}(k_v)$  &  $\propto[k_{v}D_z^{\frac{1}{2}}(S)]^{-\frac{2(1-\gamma)}{m}}$ & $\propto[k_{v}D_z^{\frac{1}{2}}(S)]^{-\frac{2(3-\gamma)}{m}}$\\\hline
$P_{v}(k_v)$ & $\propto[k_{v}D_z^{\frac{1}{2}}(S)]^{-\frac{2}{m}}$ & $\propto[k_{v}D_z^{\frac{1}{2}}(S)]^{-\frac{6}{m}}$\\
\hline
\end{tabular}
\caption{VCS near a scale $k_v$ depends on whether the instrument with beam width $\Delta B$ resolves the correspondent spatial scale $[k_v^2D_z(S)]^{-\frac{1}{m}}$, where S is the scale where turbulence saturates. $D_z^{\frac{1}{2}}(S)$ is the characteristic turbulent velocity difference at separation S. $\gamma$ is the reduced density power spectrum slope while $m$ is the velocity structure function index. $k_v$ is velocity wave number, reciprocal to
$v/D_z^{\frac{1}{2}}(S)$. P$_p(k_v)$ is the power spectrum of emissivity (proportional to density in the case of \ion{H}{1} observations) along the velocity axis, while P$_v(k_v)$ is the power spectrum of velocity fluctuations.}
\end{table*}

\begin{table*}
\centering
\begin{tabular}{| c |c | c | c | c | c | c | c | c |}
\hline
\# & Object & Reference & Data & P$_{PPV}^{thin}$ & P$_{PPV}^{thick}$ & Depth & E$_{v}$  & E$_\rho$ \\ \hline \hline
1 & Arm & \citet{2006ApJS..165..512K} & \ion{H}{1} & -2.6  & -3.4 & Thin & -1.8 &  -1.2 \\
2 & SMC & \citet{2001ApJ...551L..53S} & \ion{H}{1} & -2.7  & -3.4 & Thin & -1.7 &  -1.4 \\
3 & CygA & \citet{2000ApJ...543..227D} & \ion{H}{1} & -2.7  & -2.8 & Thin & N/A  &  -0.8\\
4 & Anticente & \citet{1993MNRAS.262..327G} & \ion{H}{1} & -2.7  & N/A & Thin & -1.7 &  -1.0 \\
5 & NGC 2592-2594 & \citet{2019MNRAS.483.3437C} & \ion{H}{1} & -2.9  & -3.1 & Thin &  N/A &  -1.1 \\
6 & L1512 & \citet{1998A...336..697S} & $^{12}$CO & N/A  & -2.8 & Thick & N/A &  -0.8 \\
7 & L1512 & \citet{1998A...336..697S} & $^{13}$CO & N/A  & -2.8 & Thick & N/A &  -0.8 \\
8 & Perseus & \citet{2006A...451..539S} & $^{13}$CO & -2.7  & -3.0 & Thick & -1.7 &  -1.0 \\
9 & Perseus & \citet{2006ApJ...653L.125P} & $^{13}$CO & -2.6  & -3.0 & Thick & -1.8  & -1.0\\
10 & L1551 & \citet{2008ApJS..174..202S} & C$^{18}$O & -2.7  & -2.8 & Thin & -1.7 &  -0.8 \\
11 & G0.253+0.016 & \citet{ 2015ApJ...802..125R} & HCN & N/A  & -3.0 & Thick & N/A &  -1.0 \\
12 & G0.253+0.016 & \citet{ 2015ApJ...802..125R} & HCO$^{+}$ & N/A  & -2.9 & Thick & N/A &  -0.9 \\
13 & G0.253+0.016 & \citet{ 2015ApJ...802..125R} & SiO & N/A  & -3.1 & Thick & N/A &  -1.1 \\
14 & Orion Nebula & \citet{2016MNRAS.463.2864A} & [S II] $\lambda$6716 & -2.7  & -3.0 & Thin & -1.6 &  -1.0 \\
15 & Orion Nebula & \citet{2016MNRAS.463.2864A} & [S II] $\lambda$6731 & -2.7  & -3.0 & Thin & -1.6 &  -1.0 \\
16 & Orion Nebula & \citet{2016MNRAS.463.2864A} & [N II] $\lambda$6583 & -2.3  & -2.6 & Thin & -1.6 &  -0.6 \\
17 & Orion Nebula & \citet{2016MNRAS.463.2864A} & H$\alpha \lambda$6563 & -2.7  & -2.8 & Thin & N/A &  -0.8 \\
18 & Orion Nebula & \citet{2016MNRAS.463.2864A} & [O III] $\lambda$5007 & -2.5 & -2.8 & Thin & -1.6 &  -0.8 \\
19 & Orion Nebula & \citet{2016MNRAS.463.2864A} &  [O III] $\lambda$5007H & -2.2 & -2.4 & Thin & -1.4 &  -0.4 \\
\hline
\end{tabular}
\caption{\label{tab:4}  P$_{PPV}^{thin}$ and  P$_{PPV}^{thick}$ are the power law spectrum in thin and thick PPV slices, respectively. $E_\rho$ is the index of density spectrum, $E_v$ is velocity spectrum. The calculation is performed by VCA. We remove the spectrum in the presence of self-absorption.}
\end{table*}

\begin{table}
\centering
\begin{tabular}{| c |c | c | c | c | }
\hline
\# & Object & Reference & Data & E$_{v}$ \\ \hline \hline
1 & SMC & \citet{2015ApJ...810...33C}  & \ion{H}{1}  & -1.85 \\
2 & HLGR & \citet{2010ApJ...714.1398C}  & \ion{H}{1}  & -1.87 \\
3 & NGC 1333 & \citet{2009ApJ...707L.153P} & $^{13}$CO  & -1.85\\
4 & NGC 6334 & \citet{2018ApJ...862...42T} & HCN  & -1.66\\
5 & NGC 6334 & \citet{2018ApJ...862...42T} & HCO$^+$  & -2.01\\
\hline
\end{tabular}
\caption{\label{tab:5}   $E_v$ is the index of velocity spectrum. The calculation is performed by VCS. HLGR represents one High-Latitude Galactic Region.}
\end{table}

All in all, in this section we showed that the analysis of the data agrees well with the predictions in LP00 and contradicts to the explanation of the change of the spectral index with the PPV slice provided in CX19.

\section{Additional points}\label{sec:dis}

\subsection{Importance of density for $M_s>1$}

For supersonic turbulence,CX19 suggests that the density spectrum gets shallow and the density fluctuations dominate. First of all, this way of reasoning is not correct as the density having shallow spectrum {\it does not suggest} that velocity fluctuations are not important for creating structures in the thin channel maps. In fact, for a shallow spectrum of density LP00 predicts that both density and velocity fluctuations are responsible for the fluctuations observed in thin channel maps.

Moreover, $M_s>1$ does not mean that the density spectrum must be shallow. in reality, MHD numerical simulations \citep{2005ApJ...624L..93B,2007ApJ...658..423K} as well as hydrodynamic numerical simulations (Kim \& Ryu 2006) testify that supersonic turbulence does not necessarily produce a shallow spectrum with $n<3$. In fact, the turbulence index is gradually changes with the sonic Mach number $M_s$ and for isothermal simulations gets shallow only for $M_s\gg 1$. In our numerical simulations we only see significant spectral flattening when $M_s>6$. In any case, a quick inspection of Tab.~\ref{tab:4} and Tab.~\ref{tab:5} shows that the density spectra of \ion{H}{1} in most cases are steep. This suggests according to LP00 the dominance of velocity fluctuations in thin channel maps. 

\subsection{Does thermal broadening remove all the velocity effects?}
\label{ssec:thermal}
One of the most important assumptions from CX19 is that, in the case of subsonic media the thermal broadening washes out the contributions from velocities and left with the density structures\footnote{CX19 Fig 2 caption: "{\it For subsonic flows, the density and velocity fields are not well correlated, but thermal broadening washes out intensity fluctuations coupled to the turbulent velocity field, and thermodynamically correlates the channel map structure with the column density map.}"}. Thus CX19 assumes that in the subsonic media the intensity structures seen in velocity channels are completely density-like.

\begin{figure}
\centering
\includegraphics[width=0.99\linewidth]{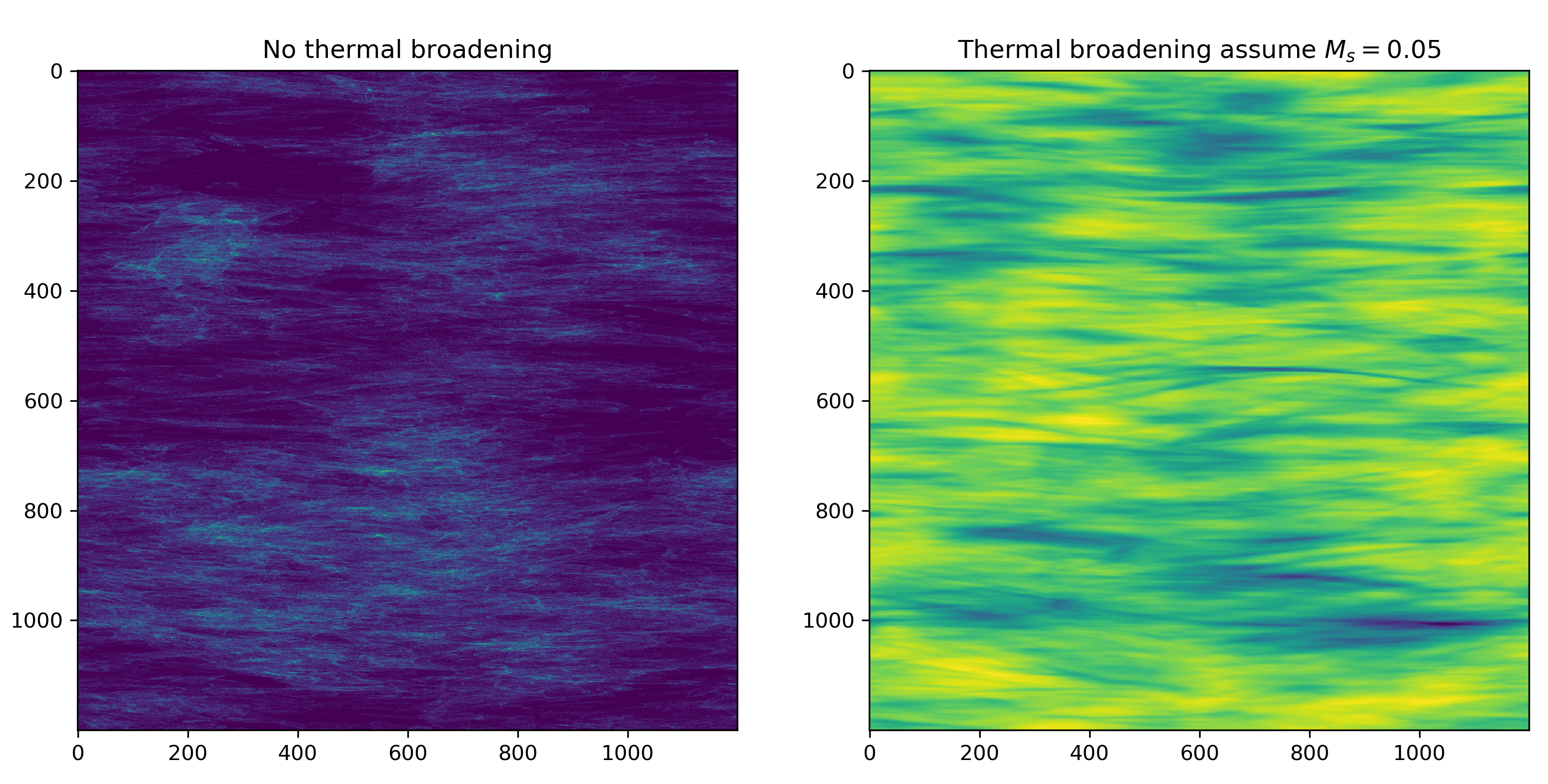}
\caption{A numerical test on whether the velocity caustics (i.e. constant density for the entire region) will show structures in the case of thermal broadening. We here {\it increase} the thermal speed to be 20 times the velocity dispersion in producing the synthetic observational maps. According to CX19 in this scenario the density map (which is a constant) should dominate. However we do see a significant structure aroused by the turbulent velocity.}
\label{fig:thermal_test}
\end{figure}

To test this claim we perform a numerical test assuming (1) density is constant in the cube (2) the thermal broadening effect is much stronger than turbulence. Fig. \ref{fig:thermal_test} shows our synthetic thin channel maps in a numerical simulation {\it with magnetic field} without (left) and with(right) thermal broadening. One can obviously see that the structures have not vanished in the case even for $M_s=0.05$. Therefore the claim that the velocity structures are washed out completely in subsonic cases does not correctly represent the reality..

\subsection{The change of LP00 in the case of temperature varying media?}

One criticism of LP00 from CX19 is related to thermal effects.
While interaction of thermal broadening with velocity channel analysis was understood and accounted for in LP00, LP00
indeed assumed that temperature is constant in space, or at least that the medium consists of phases with distinctly different constant temperatures. Effect of the spatially variable and
correlated temperature $\beta_T(\mathbf{x})$ awaits further
detailed study.  Here we provide the argument why with temperature varying case LP00's qualitative picture is still applicable.

As we stress, the most important observation by LP00 is about the change of the spectral slopes for velocity channels with different integration width along the velocity axis and \textit{that the slope depends on the spatial scale of the measurements}. Introducing temperature inhomogeneties $\beta_T(\mathbf{x})$ in Eq.~(\ref{eq:rho_PPV}) as a spatially correlated random field depending on the position would actually increase the role of the velocity effects in velocity channels since thermal effects are \it{part of the velocity modification of the PPV density}.
One can presumably treat
temperature as a passive scalar dragged by turbulent motions. Such "entropy" fluctuations will provide additional {\it positive} contribution to fluctuations in observed intensity, beyond pure PPP density inhomogeneities, and may give new information about the turbulence. However, that does not alter the fact that spectral
properties of intensity fluctuation will change with the change
of channel thickness.  As a result, the existence of temperature inhomogeneties not only increases the weight of velocity-related contributions to the velocity channel map, but also keeps the phenomenon of spectral slope transitions intact. 

\subsection{What are the limitations of LP00 theory?}

While we feel that parts of CX19 misrepresents LP00 theory and claims its limitations and problems where they do not exist, the theory does have its assumptions that potentially limit its applicability. One of them is that within LP00 derivations it is assumed that the density and velocity fluctuations are not correlated. As the benefit LP00 did not have to assume anything about statistics of the density field. However, in the Appendix D of LP00 it was demonstrated analytically on the example of log-normal model for density distribution that in the limiting case of {\it maximal possible} correlation between the density and velocity field, it is velocity effects that are amplified  in spectra from thin channels.  This seems to strongly suggest
that density velocity correlations, if anything, enforce LP00 prediction that in thin channels one can measure turbulent velocity properties. \footnote{It is important to stress that obtaining the spectrum of density does not rely on the LP00 theory.} 

The LP00 theory was formulated for the isothermal gas, which potentially can introduce uncertainties in obtaining the spectral index of velocity fluctuations. While for molecular gases, e.g. CO, the assumption of one temperature is a good one, the situation is somewhat more complicated in HI. The model of two phase media was considered in LP00 and it was claimed there that the model works if Cold phase \ion{H}{1} can be treated as passive scalar moved by the Warm phase of \ion{H}{1}. Such behavior is generally known for entropy (temperature) in fluctuations \citep{2003matu.book.....B} and is expected to be applicable to high galactic latitudes where the Cold phase of \ion{H}{1} is known to be subdominant in terms of its mass. \footnote{Note that in this model the thermal broadening is coming from the matter within the cold gas
which dominate channel intensity, while the dynamics is driven by the warm gas. This is similar to adding heavy molecules inside the flow. } 

The way of reasoning above may be generalized for the unstable \ion{H}{1} at the intermediate temperatures. The correspondence of the velocity spectral indexes obtained from observations using the LP00 based analysis with the expectations for the MHD turbulent spectra expectations (see Tab.~\ref{tab:3},\ref{tab:4},\ref{tab:5}) and the velocity spectral indexes obtained with the other techniques, e.g. velocity centroid one (see Tab.~\ref{tab:4}) support the aforementioned model of HI dynamics. The corresponding discussion is, however, beyond the scope of our comment. In terms of our present discussion, the complications related to the variations of the temperature are expected to change the velocity spectral indexes (which we do not see) rather than the general qualitative behavior of the velocity and density fluctuations that we discuss in this Comment.

\section{Summary}\label{sec:concl}
CX19 uses both observational and numerical data to address important questions related to the nature of intensity structures in PPV space. In doing so the authors question the applicability of LP00 theory to studying spectra of turbulent velocity in galactic \ion{H}{1}.  In the Comment we re-analyze the observational data used in CX19, employ turbulence simulations with significantly higher numerical resolution and come to the conclusion that all the available data is consistent with the predictions of LP00 theory as well as the expectations of the modern theory of compressible MHD turbulence. In particular:
\begin{enumerate}
    \item From the theoretical point of view, thin velocity channels are mixtures of densities and velocities. The relative importance are determined by the steepness of the spectrum, which is shown both analytically and observationally (\S \ref{sec:LP00}). The claim in CX19 that the observed intensity structure of the PPV cubes is determined either by velocity or density fluctuations inaccurate and misleading.
    \item It is  possible to have spatial correlation between thin channel maps with column density map, but actually velocity caustics contributes more within the channel (\S\ref{sec:LP00}).
    \item We examine the same observation region in GALFA which that the \ion{H}{1} structure is "shown" to be density structures with the biased parameter (\S \ref{sec:result}). 
    \item We provide an investigation using the prediction of spectral slope (\S \ref{sec:result}) to show that the region does in fact show the spectral change.
\end{enumerate}

\section*{Acknowledgment}  We acknowledge the Galactic ALFA HI (GALFA-HI) survey providing data set obtained with the Arecibo L-band Feed Array (ALFA) on the Arecibo 305m telescope.

\appendix
\section{Main results of Velocity Channel Analysis}
\label{apsec:A}
Velocity Channel Analysis (VCA) is the statistical analysis of the velocity channel data of the optically thin (\cite{LP00}) spectral lines and, later (\cite{LP04}) lines with self-absorption. The goal of VCA is to determine the statistical properties of the turbulence such as spatial density and turbulent motion in the medium via correlation (or equivalently power spectrum) analysis of the sky intensity maps in velocity channels. The focus of VCA is to find out what should be measured to disentangle density and velocity information. The possibility of such disentanglement stems from velocity sensitivity of the number of emitters (and thus intensity) in thin velocity channels, as per Eq.~\ref{eq:rho_PPV}. The main points of the VCA are:

a)
In \textit{thick} channels, the velocity effects on the intensity are integrated out, while in \textit{thin} channels they are retained. Importantly, the criterion that the channel is \textit{thin} or \textit{thick}
\begin{eqnarray}
\mathit{thin} &:& \Delta v^2 + 2 \beta_T \ll D_z(R) \\
\mathit{thick} &:& \Delta v^2 + 2 \beta_T \gg D_z(R)
\end{eqnarray}
depends on the scale of line-of-sight separation $R$ at which sky correlation function or power spectrum is measured. \\
\linebreak

b) For optically thin lines, \cite{LP00} showed that measuring the correlation function at different scales reflect different information about the turbulence. At scales sufficiently small that the channel is \textit{thick}, intensity correlation reveals the underlying density correlation. At larger scales, velocity statistics can be obtained if the \textit{thin} channel regime is achieved. In terms of the 2D power spectrum of intensity 
\begin{eqnarray}
\textit{thick} & \quad K \gg K_* &  \quad\quad P(K) \sim K^{n} \\
\textit{thin}  & \quad K \ll K_* & \quad\quad P(K) \sim \left\{
\begin{array}{ll}
K^{n+m/2} & n > -3\\
K^{-3+m/2} & n \le -3
\end{array}
\right.
\end{eqnarray}
where $n$ is the slope of the density power spectrum and $m$ is the slope
of velocity structure function ($m=2/3$ for Kolmogorov turbulence). 
The \textit{thin}--\textit{thick} transition scale $K_*$ carries information about the magnitude of the turbulent velocities.  The analysis benefits from our ability to vary the channel thickness synthetically. 

For observations performed with a finite effective slice thickness in the transitional scale range around the corresponding $K_*$, the local slope of 2D intensity power spectrum will be in-between the values given by \it{thin} and \it{thick} asymptotics. To model this behaviour accurately one needs to return to quantitative evaluation of full theoretical integral expressions for the power spectrum.

c) The dichotomy that morphological structures in velocity channels are either density or velocity, as posed by \cite{susan19}, is not the position taken in VCA papers, although it was pointed out that even for the uniform spatial density, PPV density slices will exhibit inhomogeneities (obviously purely due to velocity mapping in this case). What VCA stated is that there is a range of \textit{intermediate scales}, over which intensity correlations will be affected both by density and velocity, as given by approximate formula
\begin{equation}
    \xi_{I} ({\bf R}) \propto \int dz \frac{{\bar \rho}^2+\xi_{\delta\rho}({\bf r})}{\sqrt{D_{z}({\bf r})+2 \beta_T}}
    \label{eq:CF_simplified}
\end{equation}
that contains contributions both from overdensity $\xi_{\delta\rho}(\mathbf{r}) =\langle \delta \rho ({\bf r'})\delta \rho ({\bf r'}+{\bf r})\rangle_{\bf r'}$ with $\delta \rho = \rho - {\bar \rho}$ and velocity $D_z(\mathbf{r}) = \langle (v_z({\bf r'}-v_z({\bf r'}+{\bf r})^2 \rangle$. In the case of steep density spectrum, the corresponding intermediate power spectrum will be determined by velocity scaling.  This range is determined by the properties of turbulent medium and is limited on small scale side by the balance between magnitude of turbulent velocities and effective channel width (with temperature contribution) and on large scales is limited by the shearing action of  coherent motions over observed volume, if they are present, or energy injection scale of the turbulence. If this range exists, it can be used with accurate correlation analysis to determine velocity information from.


\begin{thebibliography}{}
\providecommand\natexlab[1]{#1}
\providecommand\JournalTitle[1]{#1}
\bibitem[(Arnold et al. 1985)]{arnold1985singularities}
  Arnold, V.I., Varchenko, A. and Gusein-Zade, S.M., Singularities of Differentiable Maps: Volume I: The Classification of Critical Points Caustics and Wave Fronts,
  1985,  Birkh{\"a}user, Boston
\bibitem[Arthur et al.(2016)]{2016MNRAS.463.2864A} Arthur, S.~J., Medina, S.-N.~X., \& Henney, W.~J.\ 2016, \mnras, 463, 2864 
\bibitem[Brandenburg \& Lazarian(2013)]{BL13} Brandenburg, A., \& Lazarian, A.\ 2013, \ssr, 178, 163 
\bibitem[Beresnyak et al.(2005)]{2005ApJ...624L..93B} Beresnyak, A., Lazarian, A., \& Cho, J.\ 2005, \apjl, 624, L93 
\bibitem[Beresnyak \& Lazarian(2015)]{2015ASSL..407..163B} Beresnyak, A., \& Lazarian, A.\ 2015, Magnetic Fields in Diffuse Media, 407, 163 
\bibitem[Burkhart et al.(2013)]{2013ApJ...771..123B} Burkhart, B., Lazarian, A., Ossenkopf, V., \& Stutzki, J.\ 2013, \apj, 771, 123 
\bibitem[Begum et al.(2006)]{2006MNRAS.372L..33B} Begum, A., Chengalur, J.~N., \& Bhardwaj, S.\ 2006, \mnras, 372, L33 
\bibitem[Biskamp(2003)]{2003matu.book.....B} Biskamp, D.\ 2003, Magnetohydrodynamic Turbulence, by Dieter Biskamp, pp.~310.~ISBN 0521810116.~Cambridge, UK: Cambridge University Press, September 2003., 310 \bibitem[Chepurnov \& Lazarian(2006)]{2006astro.ph.11465C} Chepurnov, A., \& Lazarian, A.\ 2006, arXiv:astro-ph/0611465 
\bibitem[Chrupnov, \& Lazarian(2008)]{2008arXiv0811.0845C} Chrupnov, A., \& Lazarian, A.\ 2008, arXiv e-prints , arXiv:0811.0845.
\bibitem[Chepurnov et al.(2010)]{2010ApJ...714.1398C} Chepurnov, A., Lazarian, A., Stanimirovi{\'c}, S., Heiles, C., \& Peek, J.~E.~G.\ 2010, \apj, 714, 1398 
\bibitem[Chepurnov et al.(2015)]{2015ApJ...810...33C} Chepurnov, A., Burkhart, B., Lazarian, A., et al.\ 2015, \apj, 810, 33
\bibitem[Cho \& Lazarian(2002)]{CL02} Cho, J., \& Lazarian, A.\ 2002, Physical Review Letters, 88, 245001 
\bibitem[Cho \& Lazarian(2003)]{Cho2003CompressibleImplications} Cho, J., \& Lazarian, A. 2003, \href{http://dx.doi.org/10.1046/j.1365-8711.2003.06941.x}{\JournalTitle{Monthly Notices of the Royal Astronomical Society}, 345, 325}
\bibitem[Clark et al.(2015)]{2015PhRvL.115x1302C} Clark, S.~E., Hill, J.~C., Peek, J.~E.~G., Putman, M.~E., \& Babler, B.~L.\ 2015, Physical Review Letters, 115, 241302 \bibitem[Clark et al.(2014)]{2014ApJ...789...82C} Clark, S.~E., Peek, J.~E.~G., \& Putman, M.~E.\ 2014, \apj, 789, 82.
\bibitem[Clark et al.(2019)]{susan19}  Clark, S.~E., Peek, J.~E.~G., \& Miville-Deschenes, M.-A.\ 2019, \apj, in press,  arXiv:1902.01409 {\bf (CX19)}
\bibitem[Choudhuri, \& Roy(2019)]{2019MNRAS.483.3437C} Choudhuri, S., \& Roy, N.\ 2019, \mnras, 483, 3437.
\bibitem[Deshpande et al.(2000)]{2000ApJ...543..227D} Deshpande, A.~A., Dwarakanath, K.~S., \& Goss, W.~M.\ 2000, \apj, 543, 227 
\bibitem[Dickey et al.(2001)]{2001ApJ...561..264D} Dickey, J.~M., McClure-Griffiths, N.~M., Stanimirovi{\'c}, S., Gaensler, B.~M., \& Green, A.~J.\ 2001, \apj, 561, 264 
\bibitem[Esquivel et al.(2005)]{2005AAS...206.3813E} Esquivel, A., Benjamin, R.~A., Cho, J., Lazarian, A., \& Leitner, S.~N.\ 2005, Bulletin of the American Astronomical Society, 37, 38.13 
\bibitem[Federrath \& Klessen(2012)]{2012ApJ...761..156F} Federrath, C., \& Klessen, R.~S.\ 2012, \apj, 761, 156 
\bibitem[Green(1993)]{1993MNRAS.262..327G} Green, D.~A.\ 1993, \mnras, 262, 327 
\bibitem[Goldreich \& Sridhar(1995)]{GS95} Goldreich, P., \& Sridhar, S. 1995, \href{http://dx.doi.org/10.1086/174600}{\JournalTitle{The Astronomical Journal}, 438, 763}
\bibitem[Hayes et al.(2006)]{2006ApJS..165..188H} Hayes, J.~C., Norman, M.~L., Fiedler, R.~A., et al.\ 2006, \apjs, 165, 188 
\bibitem[Hu et al.(2018)]{2018MNRAS.480.1333H} Hu, Y., Yuen, K.~H., \& Lazarian, A.\ 2018, \mnras, 480, 1333 
\bibitem[Hu et al.(2019a)]{survey} Hu, Y., Yuen, K. H.,  Lazarian V., Ho, K.W.,  Benjamin, A.R., Hill, S.A., Lockman, J.F., Goldsmith, F.P. \& Lazarian, A.,\ 2018 submitted to Nature Astronomy
\bibitem[Heiles(2001)]{2001ApJ...551L.105H} Heiles, C.\ 2001, \apjl, 551, L105 
\bibitem[Kowal et al.(2007)]{2007ApJ...658..423K} Kowal, G., Lazarian, A., \& Beresnyak, A.\ 2007, \apj, 658, 423 
\bibitem[Khalil et al.(2006)]{2006ApJS..165..512K} Khalil, A., Joncas, G., Nekka, F., Kestener, P., \& Arneodo, A.\ 2006, \apjs, 165, 512 
\bibitem[Kandel et al.(2016)]{KLP16} Kandel, D., Lazarian, A., \& Pogosyan, D.\ 2016, \mnras, 461, 1227 
\bibitem[Kandel et al.(2017a)]{KLP17a} Kandel, D., Lazarian, A., \& Pogosyan, D.\ 2017, \mnras, 464, 3617 
\bibitem[Kandel et al.(2017b)]{KLP17b} Kandel, D.,
Lazarian, A., \& Pogosyan, D.\ 2017, \mnras, 470, 3103 
\bibitem[Lazarian \& Pogosyan(2000)]{LP00} Lazarian, A., \& Pogosyan, D.\ 2000, \apj, 537, 720 {\bf (LP00)}
\bibitem[Lazarian \& Pogosyan(2004)]{LP04} Lazarian, A., \& Pogosyan, D.\ 2004, \apj, 616, 943 
\bibitem[Lazarian \& Pogosyan(2008)]{LP08} Lazarian, A., \& Pogosyan, D.\ 2008, \apj, 686, 350 
\bibitem[Lazarian \& Pogosyan(2006)]{LP06} Lazarian, A., \& Pogosyan, D.\ 2006, \apj, 652, 1348 
\bibitem[Lazarian \& Pogosyan(2012)]{LP12} Lazarian, A., \& Pogosyan, D.\ 2012, \apj, 747, 5 
\bibitem[Lazarian \& Pogosyan(2016)]{LP16} Lazarian, A., \& Pogosyan, D.\ 2016, \apj, 818, 178 
\bibitem[Lazarian \& Yuen(2018)]{LY18a} Lazarian, A., \& Yuen, K.~H.\ 2018, \apj, 853, 96 
\bibitem[Lazarian \& Yuen(2018b)]{LY18b} Lazarian, A., \& Yuen, K.~H.\ 2018, arXiv:1802.00028 
\bibitem[Lazarian(2006)]{2006AIPC..874..301L} Lazarian, A.\ 2006, Spectral Line Shapes: XVIII, 874, 301 
\bibitem[Lazarian, et al.(2018)]{2018ApJ...865...46L} Lazarian A., et al., 2018, ApJ, 865, 46
\bibitem[Lazarian \& Vishniac(1999)]{LV99} Lazarian, A., \& Vishniac, E.~T.\ 1999, \apj, 517, 700 
\bibitem[Muller et al.(2004)]{2004ApJ...616..845M} Muller, E., Stanimirovi{\'c}, S., Rosolowsky, E., et al.\ 2004, \apj, 616, 845.
\bibitem[Miville-Desch{\^e}nes et al.(2003)]{2003A&A...411..109M} Miville-Desch{\^e}nes, M.-A., Joncas, G., Falgarone, E., \& Boulanger, F.\ 2003, \aap, 411, 109 
\bibitem[Nguyen et al.(2018)]{2018ApJ...862...49N} Nguyen, H., Dawson, J.~R., Miville-Desch{\^e}nes, M.-A., et al.\ 2018, \apj, 862, 49.
\bibitem[Rebolledo et al.(2017)]{2017MNRAS.472.1685R} Rebolledo, D., Green, A.~J., Burton, M., et al.\ 2017, \mnras, 472, 1685 
\bibitem[Planck Collaboration et al.(2016)]{2016A&A...586A.138P} Planck Collaboration, Ade, P.~A.~R., Aghanim, N., et al.\ 2016, \aap, 586, A138 
\bibitem[Peek et al.(2018)]{susantail} Peek, J.~E.~G., Babler, B.~L., Zheng, Y., et al.\ 2018, \apjs, 234, 2 
\bibitem[Padoan et al.(2006)]{2006ApJ...653L.125P} Padoan, P., Juvela, M., Kritsuk, A., \& Norman, M.~L.\ 2006, \apjl, 653, L125 
\bibitem[Padoan et al.(2009)]{2009ApJ...707L.153P} Padoan, P., Juvela, M., Kritsuk, A., et al.\ 2009, \apj, 707, L153.
\bibitem[Robitaille et al.(2010)]{2010MNRAS.405..638R} Robitaille, J.-F., Joncas, G., \& Khalil, A.\ 2010, \mnras, 405, 638 
\bibitem[Rathborne et al.(2015)]{2015ApJ...802..125R} Rathborne, J.~M., Longmore, S.~N., Jackson, J.~M., et al.\ 2015, \apj, 802, 125 
\bibitem[Stanimirovi{\'c} \& Lazarian(2001)]{2001ApJ...551L..53S} Stanimirovi{\'c}, S., \& Lazarian, A.\ 2001, \apjl, 551, L53 
\bibitem[Stutzki et al.(1998)]{1998A...336..697S} Stutzki, J., Bensch, F., Heithausen, A., Ossenkopf, V., \& Zielinsky, M.\ 1998, \aap, 336, 697 
\bibitem[Soler et al.(2018)]{Soler2018} Soler, J.~D., Beuther, H., Rugel, M., et al.\ 2018, arXiv:1809.08338 
\bibitem[Swift, \& Welch(2008)]{2008ApJS..174..202S} Swift, J.~J., \& Welch, W.~J.\ 2008, The Astrophysical Journal Supplement Series, 174, 202.
\bibitem[Sun et al.(2006)]{2006A...451..539S} Sun, K., Kramer, C., Ossenkopf, V., et al.\ 2006, \aap, 451, 539 
\bibitem[Tang et al.(2018)]{2018ApJ...862...42T} Tang, K.~S., Li, H.-B., \& Lee, W.-K.\ 2018, \apj, 862, 42 
\bibitem[Xu \& Lazarian(2018)]{2018arXiv180200987X} Xu, S., \& Lazarian, A.\ 2018, arXiv:1802.00987 
\bibitem[Yuen \& Lazarian(2017a)]{YL17a} Yuen, K.~H., \& Lazarian, A.\ 2017, \apjl, 837, L24 
\bibitem[Yuen, \& Lazarian(2017b)]{YL17b} Yuen, K.~H., \& Lazarian, A.\ 2017, arXiv e-prints , arXiv:1703.03026.
\bibitem[Ysard et al.(2015)]{2015A&A...577A.110Y} Ysard, N., K{\"o}hler, M., Jones, A., et al.\ 2015, \aap, 577, A110 
\bibitem[Zhang et al.(2012)]{2012ApJ...754...29Z} Zhang, H.-X., Hunter, D.~A., \& Elmegreen, B.~G.\ 2012, \apj, 754, 29 
\end{thebibliography}
\end{document}